\documentclass[proceedings]{JHEP3}

\PrHEP{PrHEP hep2001}
\conference{International Europhysics Conference on HEP}

\usepackage{epsfig}
\usepackage{amsmath}
\usepackage{amssymb}

\newcommand{\eff}{\text{eff}}
\newcommand{\Order}{{\mathcal O}}
\newcommand{\bra}{\langle}
\newcommand{\ket}{\rangle}
\newcommand{\s}{\hat{s}}
\newcommand{\bsll}{b\to s\ell^{+}\ell^{-}}
\newcommand{\Bsll}{B\to X_s\ell^{+}\ell^{-}}
\newcommand{\pole}{\text{pole}}

\title{Results of the $\boldsymbol{\Order(\!\alpha_s\!)}$ two-loop virtual
       corrections to $\boldsymbol{B \to X_s \, \ell^+ \ell^-}$ in the standard model}

\author{\speaker{Christoph Greub}                          
         and M. Walker
         \thanks{Work partially supported by Schweizerischer Nationalfonds and SCOPES program.}\\
        Institut f\"ur Theoretische Physik, Universit\"at Bern, CH--3012 Bern, Switzerland \\ 
        E-mail: \email{greub@itp.unibe.ch}, \email{walker@itp.unibe.ch}}                

\abstract{We present the results of the $\Order(\alpha_s)$ two-loop virtual corrections to the differential
          decay width $d\Gamma(B\to X_s \ell^+ \ell^-)/d\s$, where $\s$ is the invariant
          mass squared of the lepton pair, normalized to $m_b^2$.
          Those contributions from gluon bremsstrahlung which are needed to
          cancel infrared and collinear singularities are also included. Our
          calculation is restricted to the range $0.05 \le \s \le 0.25$ where the effects from
          resonances are small. The new contributions drastically reduce the renormalization scale
          dependence of existing results for $d\Gamma(B\to X_s \ell^+ \ell^-)/d\s$.
          The renormalization scale uncertainty of the corresponding branching ratio
          (restricted to $0.05 \le \s \le 0.25$) gets reduced from $\sim \pm 13\%$ to $\sim \pm 6.5\%$.
         }

\begin{document}
%
%
  \section{Introduction}
    \label{sec:intro}
    After the observation of the penguin-induced decay $B\to X_s \gamma$ \cite{CLEOrare1} and corresponding exclusive
    channels such as $B\to K^*\gamma$ \cite{CLEOrare2}, rare $B$-decays have begun to play an important role in the
    phenomenology of particle physics. They put strong constraints on various extensions of the standard model.
    The inclusive decay $\Bsll$ has not been observed so far, but is expected to be detected at the currently running
    $B$-factories.

    The next-to-leading logarithmic (NLL) result for $\Bsll$ suffers from a relatively large ($\pm16\%$) dependence on
    the matching scale $\mu_W$ \cite{Misiak:1993bc,Buras:1995dj}. The NNLL corrections to the Wilson coefficients
    remove the matching scale dependence to a large extent \cite{Bobeth:2000mk}, but leave a $\pm13\%$-dependence on the
    renormalization scale $\mu_b$, which is of $\Order(m_b)$.
    In order to further improve the result, we have recently calculated the $\Order(\alpha_s)$ two-loop corrections to
    the matrix elements of the operators $O_1$ and $O_2$ as well as the $\Order(\alpha_s)$ one-loop corrections to
    $O_7$,...,~$O_{10}$~\cite{letter}.
    Because of large resonant contributions from $\bar c c$ intermediate states, we restrict the invariant lepton mass
    squared $s$ to the region $0.05 \le \s \le 0.25$, where $\s = s/m_b^2$. In the following we present a summary of
    the results of these calculations.
%
%
  \section{Theoretical Framework}
    \label{sec:framework}
    The appropriate tool for studies on weak $B$-mesons decays is the effective Hamiltonian technique. The effective
    Hamiltonian is derived from the standard model by integrating out the $t$-quark, the $Z_0-$ and the $W$-boson.
    For the decay channels $b\to s \ell^+\ell^-$ ($\ell=\mu,e$) it reads
    \begin{equation*}
        \label{Heff}
        {\cal H}_{\eff} =  - \frac{4 \, G_F}{\sqrt{2}} \, V_{ts}^* \, V_{tb} \sum_{i=1}^{10} C_i \, O_i \, ,
    \end{equation*}
    where $O_i$ are dimension six operators and $C_i$ denote the corresponding Wilson coefficients.
    The operators can be chosen as \cite{Bobeth:2000mk}
    \begin{equation*}
    \label{oper}
    \begin{array}{rclrcl}
        O_1    & = & (\bar{s}_{L}\gamma_{\mu} T^a c_{L })
                    (\bar{c}_{L }\gamma^{\mu} T^a b_{L}) &
        O_2    & = & (\bar{s}_{L}\gamma_{\mu}  c_{L })
                    (\bar{c}_{L }\gamma^{\mu} b_{L}) \\ \vspace{0.2cm}
        O_3    & = & (\bar{s}_{L}\gamma_{\mu}  b_{L })
                    \sum_q (\bar{q}\gamma^{\mu}  q) &
        O_4    & = & (\bar{s}_{L}\gamma_{\mu} T^a b_{L })
                    \sum_q (\bar{q}\gamma^{\mu} T^a q) \\ \vspace{0.2cm}
        O_5    & = & (\bar{s}_L \gamma_{\mu} \gamma_{\nu}
                    \gamma_{\sigma}b_L)
                    \sum_q(\bar{q} \gamma^{\mu} \gamma^{\nu}\gamma^{\sigma}q) &
        O_6    & = & (\bar{s}_L \gamma_{\mu} \gamma_{\nu}
                    \gamma_{\sigma} T^a b_L)
                    \sum_q(\bar{q} \gamma^{\mu} \gamma^{\nu}
                    \gamma^{\sigma} T^a q)    \vspace{0.2cm} \\ \vspace{0.2cm}
        O_7    & = & \frac{e}{g_s^2} m_b (\bar{s}_{L} \sigma^{\mu\nu}
                    b_{R}) F_{\mu\nu} &
        O_8    & = & \frac{1}{g_s} m_b (\bar{s}_{L} \sigma^{\mu\nu}
                    T^a b_{R}) G_{\mu\nu}^a \\ \vspace{0.2cm}
        O_9    & = & \frac{e^2}{g_s^2}(\bar{s}_L\gamma_{\mu} b_L)
                    \sum_\ell(\bar{\ell}\gamma^{\mu}\ell) &
        O_{10} & = & \frac{e^2}{g_s^2}(\bar{s}_L\gamma_{\mu} b_L)
                    \sum_\ell(\bar{\ell}\gamma^{\mu} \gamma_{5} \ell) \, .
    \end{array}
    \end{equation*}
    The subscripts $L$ and $R$ refer to left- and right- handed fermion fields. We work in the approximation
    where the combination $(V_{us}^* V_{ub})$ of Cabibbo-Kobayashi-Maskawa (CKM) matrix elements is neglected.
    The CKM structure factorizes therefore.
%
%
  \section{Virtual Corrections to the Operators $\boldsymbol{O_1,~O_2,~O_7,~O_8,~O_9}$ and $\boldsymbol{O_{10}}$}
  \label{sec:virt}
    Using the naive dimensional regularization scheme in $d=4-2\,\epsilon$ dimensions, ultraviolet and infrared
    singularities both show up as $1/\epsilon^n$-poles ($n=1,2$).
    The ultraviolet singularities cancel after including the counterterms. Collinear singularities are regularized by
    retaining a finite strange quark mass $m_s$. They are cancelled together with the infrared singularities at the level
    of the decay width, when taking the bremsstrahlung process $b \to s \ell^+\ell^- g$ into account.
    Gauge invariance implies that the QCD-corrected matrix elements of the operators $O_i$ can be written as
    \begin{equation*}
        \label{formdef}
            \bra s\ell^+\ell^-|O_i|b\ket =
                \hat{F}_i^{(9)} \bra O_9 \ket_{\text{tree}} +
                \hat{F}_i^{(7)} \bra O_7 \ket_{\text{tree}} \, ,
    \end{equation*}
    where  $\bra O_9 \ket_{\text{tree}}$ and  $\bra O_7 \ket_{\text{tree}}$ are the tree-level matrix elements of
    $O_9$ and $O_7$, respectively.

    \subsection{Virtual corrections to $\boldsymbol{O_1}$ and $\boldsymbol{O_2}$}
    \label{subsec:virtO12}
    For the calculation of the two-loop diagrams associated with $O_1$ and $O_2$ we mainly used a combination of
    Mellin-Barnes technique \cite{letter,Greub:2001sy} and of Taylor series expansion in $s$.
    For $s < m_b^2$ and $s < 4\, m_c^2$, most diagrams allow the latter. The unrenormalized form factors
    $\hat{F}^{(7,9)}$ of $O_1$ and $O_2$ are then obtained in the form
    \begin{equation*}
        \hat{F}^{(7,9)} = \sum_{i,j,l,m} c_{ijlm}^{(7,9)} \, \s^i \, \ln^j\!(\s) \,
                          \left(\hat{m}_c^2\right)^l \ln^m\!\left( \hat{m}_c\right),
    \end{equation*}
    where $\hat{m}_c=\frac{m_c}{m_b}$. The indices $i,j,m$ are non-negative integers
    and $l=-i,-i+\frac{1}{2},-i+1,....\,$.

    Besides the counterterms from quark field, quark mass and coupling constant ($g_s$) renormalization, there are
    counterterms induced by operator mixing. They are of the form
    \begin{equation*}
        C_i \cdot \sum_j \delta Z_{ij}\bra O_j\ket \quad \text{with} \quad
        \delta Z_{ij} = \frac{\alpha_s}{4 \, \pi} \left[ a_{ij}^{01} + \frac{a_{ij}^{11}}{\epsilon} \right] +
        \frac{\alpha_s^2}{(4 \, \pi)^2} \left[ a_{ij}^{02} + \frac{a_{ij}^{12}}{\epsilon} +
        \frac{a_{ij}^{22}}{\epsilon^2}\right] + \Order\!\left(\alpha_s^3\right).
    \end{equation*}
    A complete list of the coefficients $a_{ij}^{lm}$ used for our calculation can be found in \cite{letter}.
    The operator mixing involves also the evanescent operators
    \begin{align*}
        O_{11} & = \left( \bar s_L \gamma_{\mu} \gamma_{\nu} \gamma_{\sigma} T^a c_L \right)
                 \left( \bar c_L \gamma^{\mu} \gamma^{\nu} \gamma^{\sigma} T^a b_L \right) - 16 \, O_1
        \quad \text{and}\\
        O_{12} & = \left( \bar s_L \gamma_{\mu} \gamma_{\nu} \gamma_{\sigma} c_L \right)
                 \left( \bar c_L \gamma^{\mu} \gamma^{\nu} \gamma^{\sigma} b_L \right) - 16 \, O_2\,.
    \end{align*}

    \subsection{Virtual corrections to $\boldsymbol{O_7}$, $\boldsymbol{O_8}$, $\boldsymbol{O_9}$ and $\boldsymbol{O_{10}}$}
    \label{subsec:virtO789}
    The renormalized contributions from the operators $O_7$, $O_8$ and $O_9$ can all be written in the form
    \begin{equation*}
        \bra s \ell^+ \ell^-|C_i O_i | b \ket =
            \widetilde{C}_i^{(0)} \left( -\frac{\alpha_s}{4 \pi} \right)
            \left[ F_i^{(9)} \bra \widetilde{O}_9 \ket_{\text{tree}} +
                   F_i^{(7)} \bra \widetilde{O}_7 \ket_{\text{tree}} \right],
    \end{equation*}
    with $\widetilde{O}_i = \frac{\alpha_s}{4 \, \pi} \, O_i$\, , \quad
    $\widetilde{C}_{7,8}^{(0)} =C_{7,8}^{(1)}$\quad and \quad
    $\widetilde{C}_9^{(0)} = \frac{4 \, \pi}{\alpha_s} \left(C_9^{(0)}+\frac{\alpha_s}{4 \, \pi} C_9^{(1)}\right)$. \\
    The formally leading term $\sim g_s^{-2} C_9^{(0)}(\mu_b)$ to the amplitude for $\bsll$ is smaller than
    the NLL term $\sim g_s^{-2} [g_s^2/(16 \, \pi^2)] \, C_9^{(1)}(\mu_b)$ \cite{Grinstein:1989}. We therefore
    adapt our systematics to the numerical situation and treat the sum of these two terms as a NLL contribution, as
    indicated by the expression for $\widetilde{C}_9^{(0)}$. The decay amplitude then starts out with a NLL term.

    The contribution from $O_8$ is finite, whereas those from $O_7$ and $O_9$ are not, ie $ F_7^{(7)}$ and $ F_9^{(9)}$
    suffer from the same infrared divergent part $f_{\text{inf}}$.

    As the hadronic parts of the operators $O_9$ and $O_{10}$ are identical, the QCD corrected matrix element of $O_{10}$
    can easily be obtained from that of $O_9$.
%
%
  \section{Bremsstrahlung Corrections}
  \label{sec:brems}
    It is known \cite{Misiak:1993bc,Buras:1995dj} that the contribution to the inclusive decay width from the
    interference between the tree-level and the one-loop matrix elements of $O_9$ and  from the corresponding
    bremsstrahlung corrections can be written as
    \begin{equation*}
        \label{gamma99}
        \frac{d\Gamma_{99}}{d\hat s} = \left(\frac{\alpha_{em}}{4 \, \pi}\right)^2
        \frac{G_F^2 \, m_{b,\pole}^5 \left|V_{ts}^*V_{tb}\right|^2}{48 \, \pi^3} \, (1 - \s)^2
        \left( 1 + 2\,\s \right)
        \left[ 2\left| \widetilde C_9^{(0)} \right|^2 \frac{\alpha_s}{\pi} \, \omega_9(\s) \right].
    \end{equation*}
    Analogous formulas hold true for the contributions from $O_7$ and the interference terms between the matrix
    elements of $O_7$ and $O_9$:
    \begin{align*}
        \frac{d\Gamma_{77}}{d\hat s} & = \left(\frac{\alpha_{em}}{4 \, \pi}\right)^2
            \frac{G_F^2 \, m_{b,\pole}^5 \left|V_{ts}^*V_{tb}\right|^2}{48 \, \pi^3} \, (1 - \s)^2 \,
            4 \left( 1 + 2/\s \right)
            \left[ 2\left| \widetilde C_7^{(0)} \right|^2 \frac{\alpha_s}{\pi} \, \omega_7(\s) \right] \, , \\
        \frac{d\Gamma_{79}}{d\hat s} & = \left(\frac{\alpha_{em}}{4 \, \pi}\right)^2
            \frac{G_F^2 \, m_{b,\pole}^5 \left|V_{ts}^*V_{tb}\right|^2}{48 \, \pi^3} \, (1 - \s)^2 \,
            12 \cdot 2 \, \frac{\alpha_s}{\pi} \, \omega_{79}(\s)\,
            \mbox{Re}\!\left[ \widetilde C_7^{(0)} \widetilde C_9^{(0)}\right] \, .
    \end{align*}
    The function $\omega_9(\s) \equiv \omega(\s)$ can be found eg in in \cite{Misiak:1993bc,Buras:1995dj}.
    For $\omega_7(\s)$ and $\omega_{79}(\s)$ see \cite{letter}.
    All other bremsstrahlung corrections are finite and will be given in \cite{AAGW}.
%
%
  \section{Corrections to the Decay Width for $\boldsymbol{B \rightarrow X_s \ell^+ \ell^-}$}
  \label{sec:decaywidth}
    Combining the virtual corrections discussed in section \ref{sec:virt} with the bremsstrahlung contributions
    considered in section \ref{sec:brems}, we find for the decay width
    \begin{multline}
    \label{eq:decaywidth}
        \frac{d\Gamma(b\to X_s \ell^+\ell^-)}{d\s} =
        \left( \frac{\alpha_{em}}{4 \, \pi} \right)^2
        \frac{G_F^2 \, m_{b,\pole}^5 \left| V_{ts}^*V_{tb} \right |^2}{48 \, \pi^3} \, ( 1 - \s )^2 \times \\
        \left( \left( 1 + 2\,\s \right) \left[ \left| \widetilde C_9^{\eff} \right|^2 +
        \left| \widetilde C_{10}^{\eff} \right|^2 \right] +
        4 \left( 1 + 2/\s \right) \left| \widetilde C_7^{\eff} \right|^2 +
        12\,\, \mbox{Re}\! \left[ \widetilde C_7^{\eff}\widetilde C_9^{\eff*} \right] \right) \, ,
    \end{multline}
    where the effective Wilson coefficients $\widetilde{C}_7^{\eff}$, $\widetilde{C}_9^{\eff}$
    and $\widetilde{C}_{10}^{\eff}$ can be written as
    \begin{eqnarray}
        \label{eq:effcoeff}
        \nonumber
        \widetilde C_9^{\eff} & = &
            \left[ 1 + \frac{\alpha_s(\mu)}{\pi} \, \omega_9(\s) \right]
            \Big( A_9 + T_9 \, h(\hat m_c^2,\s) + U_9 \, h(1,\s) + W_9 \, h(0,\s) \Big)
            \\ \nonumber && \hspace{6cm}
            - \frac{\alpha_{s}(\mu)}{4\,\pi} \left( C_1^{(0)} F_1^{(9)} +
                    C_2^{(0)} F_2^{(9)} + A_8^{(0)} F_8^{(9)} \right)\, ,
        \\
        \nonumber
        \\
        \nonumber
        \widetilde C_7^{\eff} & = &
            \left[ 1 + \frac{\alpha_s(\mu)}{\pi} \, \omega_7(\s) \right] A_7
            - \frac{\alpha_{s}(\mu)}{4\,\pi} \left( C_1^{(0)} F_1^{(7)} +
                    C_2^{(0)} F_2^{(7)} + A_8^{(0)} F_8^{(7)} \right)\, ,
        \\
        \nonumber
        \widetilde C_{10}^{\eff} & = &
            \left[ 1 + \frac{\alpha_s(\mu)}{\pi} \, \omega_9(\s) \right] A_{10} \, .
    \end{eqnarray}
    The function $h(\hat m_c^2,\s)$ is defined in \cite{Bobeth:2000mk}, where also the values of $A_7$, $A_9$, $A_{10}$,
    $T_9$, $U_9$ and $W_9$ can be found. $C_1^{(0)}$, $C_2^{(0)}$ and $A_8^{(0)}=\widetilde{C}_8^{(0,\eff)}$ are taken
    from \cite{Greub:2001sy}.
%
%
  \section{Numerical Results}
  \label{sec:numres}
    The decay width in eq~(\ref{eq:decaywidth}) has a large uncertainty due to the factor $m_{b,\pole}^5$. Following common
    practice, we consider the ratio
    \begin{equation*}
        R_{\text{quark}}(\s) =
            \frac{1}{\Gamma(b \to X_c \, e \, \bar\nu_e)} \frac{d\Gamma(b \to X_s \ell^+ \ell^-)}{d\hat s} \, ,
    \end{equation*}
    in which the factor $m_{b,\pole}^5$ drops out.
    $\Gamma(b \to X_c \, e \, \bar\nu_e)$ can be found eg in \cite{Bobeth:2000mk}.

    In Fig.~\ref{fig:mudep} we investigate the dependence of $R_{\text{quark}}(\s)$ on the renormalization scale $\mu_b$
    for $0.05 \le \s \le 0.25$. The solid lines take the new NNLL contributions into account, whereas the dashed lines
    include the NLL results combined with the NNLL corrections to the matching conditions \cite{Bobeth:2000mk}, only.
    The lower, middle and upper line each correspond to $\mu_b=2.5,~5$ and 10 GeV, respectively, and $\hat m_c = 0.29$.
    From this figure we conclude that the renormalization scale dependence gets reduced by more than a factor of 2.
    For the integrated quantity we get
    \begin{equation*}
        \label{eq:Rint}
        R_{\text{quark}} = \int_{0.05}^{0.25} \, d\s \, R_{\text{quark}}(\s) = (1.25 \pm 0.08 ) \times 10^{-5} \, ,
    \end{equation*}
    where the error is obtained by varying $\mu_b$ between 2.5 GeV and 10 GeV.
    Not including our corrections, one finds $R_{\text{quark}}=(1.36 \pm 0.18)\times 10^{-5}$ \cite{Bobeth:2000mk}. In other
    words, the renormalization scale dependence got reduced from $\sim \pm 13\%$ to $\sim \pm  6.5\%$.
    The largest uncertainty due to the input parameters is induced by $\hat{m}_c$. Fig.~\ref{fig:mcdep}
    illustrates the dependence of $R_{\text{quark}}(\s)$ on $\hat{m}_c$. The dashed, solid and dash-dotted lines
    correspond to $\hat{m}_c = 0.27$, $\hat{m}_c = 0.29$ and $\hat{m}_c = 0.31$, respectively, and $\mu_b=5$ GeV.
    We find an uncertainty of $\pm 7.6\%$ due to $\hat{m}_c$.

    \DOUBLEFIGURE
        {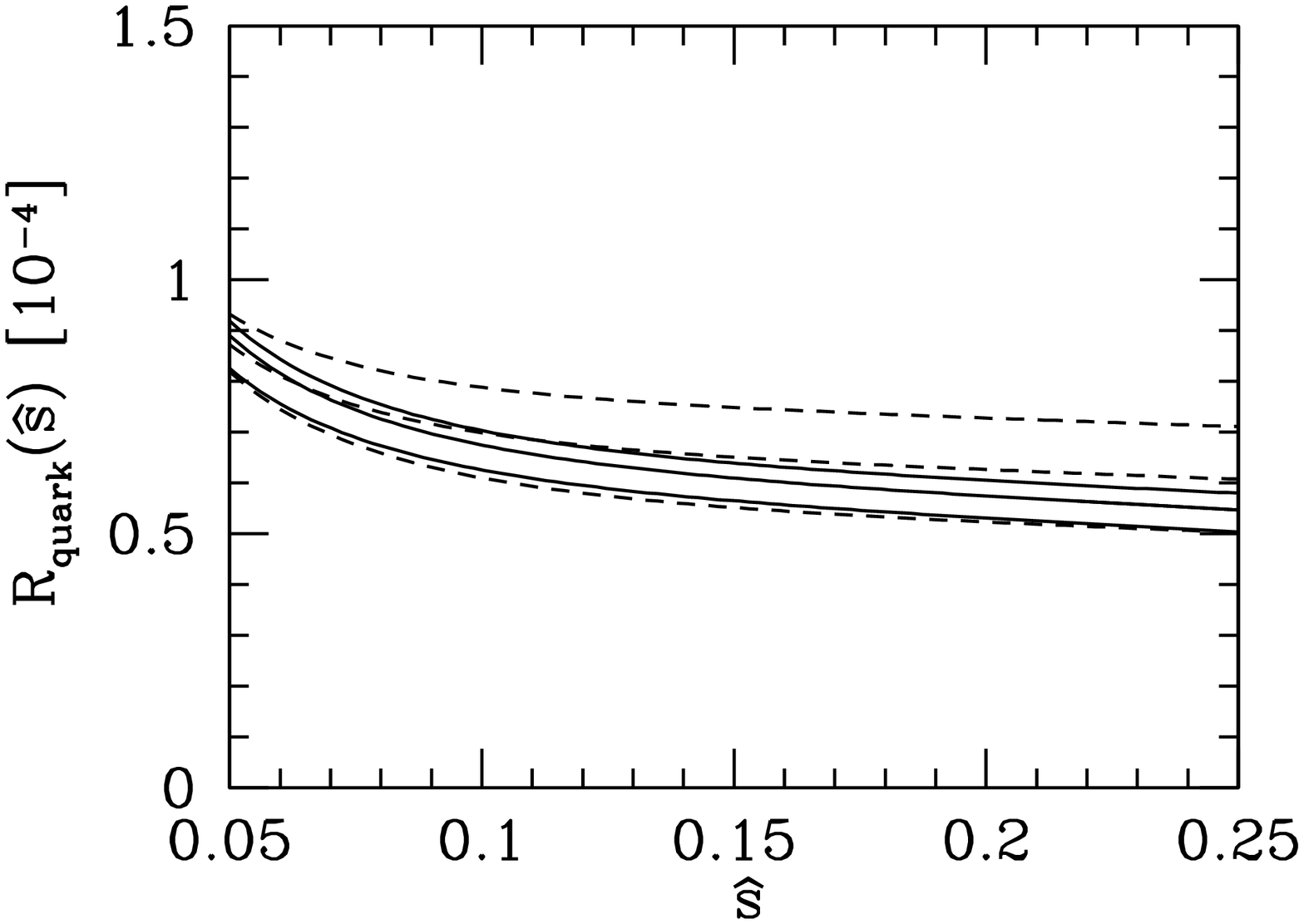,width=7cm,height=4.7cm}
        {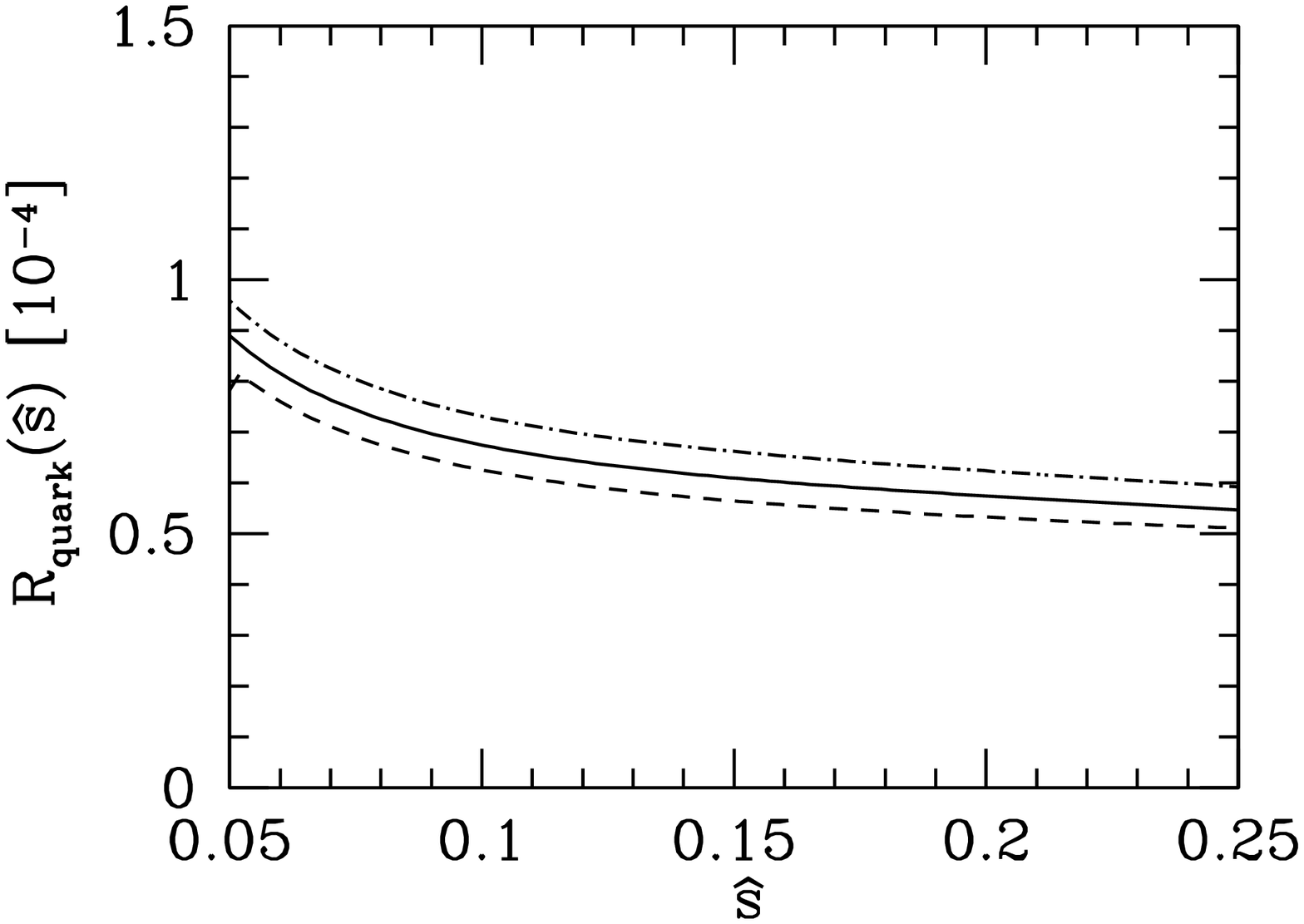,width=7cm,height=4.7cm}
        {Dependence of $R_{\text{quark}}(\s)$ on $\mu_b$.\label{fig:mudep}}
        {Dependence of $R_{\text{quark}}(\s)$ on $\hat{m}_c$.\label{fig:mcdep}}

    We conclude with the remark that the results presented in this exposition have recently been included in a systematic
    description of the corresponding exclusive decay \linebreak mode $B \to K^* \ell^+ \ell^-$ \cite{beneke, feldmann}.

%
%

\end{document}